\documentclass[10pt, conference, letterpaper]{IEEEtran}

\usepackage{amsmath}
\usepackage{array}
\usepackage{cite}
\usepackage{amsmath,amssymb,amsfonts}
\usepackage{algorithmic}
\usepackage{graphicx}
\usepackage{float}
\usepackage{booktabs}  
\usepackage{mathtools} 
\usepackage[colorlinks=true,urlcolor=blue,linkcolor=blue,citecolor=blue,bookmarks=false]{hyperref}
\usepackage{amsthm} 
\usepackage{graphicx}
\usepackage{algorithm}
\usepackage{algorithmic}

\usepackage{graphicx} 
\usepackage{subcaption}
\usepackage{comment}
\usepackage{soul}
\usepackage[usenames,dvipsnames]{xcolor}
\usepackage{svg}
\usepackage{enumitem}
\setlist[enumerate]{topsep=0pt,itemsep=-1ex,partopsep=1ex,parsep=1ex,leftmargin=4ex}
\setlist[itemize]{topsep=0pt,itemsep=-1ex,partopsep=1ex,parsep=1ex,leftmargin=4ex}

\title{Toward Practical Age-of-Information Scheduling in 5G Cellular}

\author{Zhuoyi Zhao and Igor Kadota \\
\thanks{Zhuoyi Zhao and Igor Kadota are with the Department of Electrical and Computer Engineering, Northwestern University, USA. E-mail: zhuoyizhao2025@u.northwestern.edu and kadota@northwestern.edu.}
\thanks{This paper was presented in part at~\cite{zhao2025optimizing}.}
}

\begin{document}

\maketitle

\begin{abstract}
We consider a 5G cellular network where a gNB schedules time-sensitive uplink transmissions from multiple UEs and forwards received packets to remote destinations. In practical 5G networks, the gNB does not directly observe the destination-side Age of Information (AoI) and must make scheduling decisions under stringent slot-level runtime constraints. In this paper, we develop a low-complexity AoI-aware scheduling policy for 5G cellular under limited observability. We first design a low-complexity estimator that infers UE-side packet timestamps and destination-side AoI from gNB-visible observations. Based on these estimates, we propose and implement a Max-Weight policy (MW-LC) in NetSim, a 5G emulator with a standards-compatible protocol stack, to showcase its performance against baseline 5G scheduling policies. Furthermore, we use MATLAB simulations to show that the LC estimator and MW-LC achieve performance close to a richer estimator-based AoI policy from the literature. The estimator may be of independent interest to the community, enabling AoI-aware algorithms beyond 5G scheduling. 
\end{abstract}


\vspace{-0.5\baselineskip}
\section{Introduction}\label{sec:intro}
\vspace{-1ex}
\IEEEPARstart{T}{he} Age of Information (AoI) metric has emerged as an important measure of information freshness~\cite{kaul2011minimizing,sun2017update}. Most existing AoI works are theory-oriented and highlight the promise of AoI as a design objective for communication networks~\cite{AoIsurvey}. In particular, there is a vast literature on AoI-based scheduling, e.g.,~\cite{kadota2019scheduling,gorle2025aoi,zhao2025optimizing2}, which is especially appealing for 5G cellular networks that support time-sensitive status updates~\cite{zhou2024goal}. However, a main challenge in designing practical AoI-based scheduling policies is that the 5G scheduler 
%
%
requires knowledge of two types of timestamps: (i) the generation times of the packets currently stored in the User Equipment (UE), and (ii) the generation times of the freshest packets successfully delivered to the destinations. In general, 5G Radio Access Network (RAN) nodes, referred to as gNBs, as well as other decision-makers that operate in the middle of the network, do not have direct access to UE- or destination-side timestamps~\cite{3gpp38211}.



\begin{figure}[h!]
    \centering
        \centering
        \includegraphics[width=0.38\textwidth]{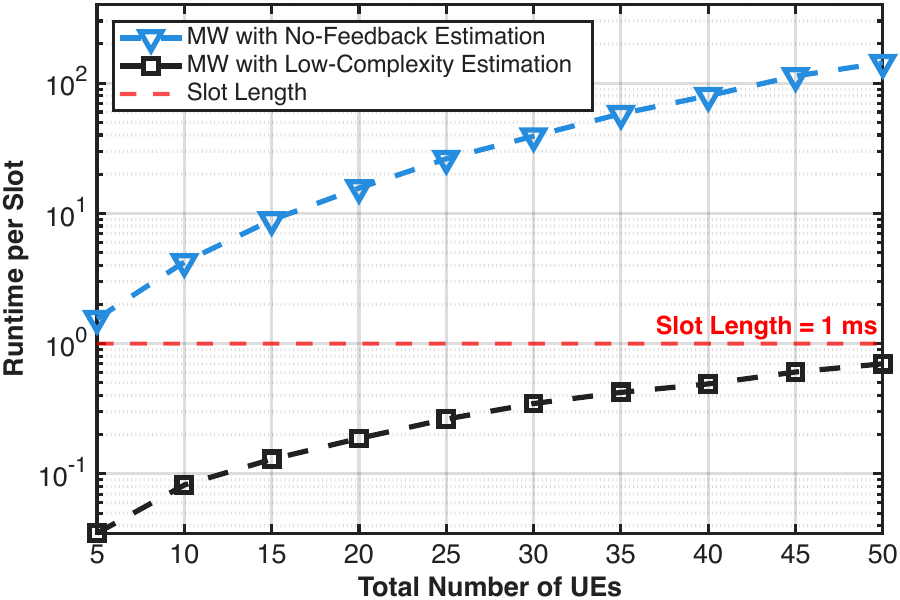}
    \caption{
    Average per-slot runtime of AoI scheduling with no-feedback estimation~\cite{zhao2025optimizing}, and with the proposed low-complexity estimation, versus the total number of users $N\in\{5,10,\ldots,50\}$. The dashed red line marks a 1\,ms TTI, which serves as a representative runtime budget for 5G scheduling. Runtime measurements were obtained in MATLAB R2025b on a machine with an Apple M2 Pro chip.}
    \vspace{-2\baselineskip}
    \label{fig:runtime_vs_users_combined}
\end{figure}
To overcome this challenge, prior system-oriented works, e.g.,~\cite{WiSwarm,AoI5G,li2024aequitas,WIFresh,ayan2021experimental,AoIOFDMA}, have taken three main directions. 
A first class of works, e.g.,~\cite{WiSwarm,AoI5G,li2024aequitas}, consider specific network settings in which AoI could be known by the gNB, for example, when UEs generate packets periodically and the AoI is measured with respect to the gNB, as opposed to the final destination. 
A second class of works, e.g.,~\cite{WIFresh,ayan2021experimental}, use Software Defined Radios to implement 
non-standard-compliant mechanisms to expose timestamp information to the scheduler. 
A third class of works optimize AoI indirectly, for example, by optimizing a grouped cyclic scheduling structure~\cite{AoIOFDMA}.

We recently proposed an estimator-based method to track the evolution of AoI over time~\cite{zhao2025optimizing}. While that work showed that AoI-aware scheduling can be realized without direct timestamp access, it has two important limitations from a 5G implementation perspective: prohibitively high computational complexity and assuming knowledge of detailed UE packet-generation statistics. 
This work shows that AoI-aware scheduling can be realized in a 5G setting when detailed packet-generation statistics and freshness are not directly observable at the gNB, by combining low-complexity estimation with scheduler design under strict runtime constraints.
Figure~\ref{fig:runtime_vs_users_combined} shows that the per-slot runtime (of estimation and scheduling) of~\cite{zhao2025optimizing} is significantly larger than the 1\'ms transmission time interval (TTI), whereas the proposed solution remains within this runtime budget across the tested system sizes, which motivates the low-complexity design pursued in this paper. 
Our main contributions can be summarized as follows:

\vspace{-0.5ex}
\begin{itemize}[leftmargin=0.15in]
\item We develop a low-complexity (LC) estimator for UE-side packet timestamps and destination-side AoI based on gNB-visible observations. Based on these estimates, we develop an AoI-aware Max-Weight scheduling policy, termed MW-LC, that incorporates uplink channel information. The resulting estimator-and-scheduler pipeline has $\mathcal{O}(N)$ per-slot complexity under sequential execution and $\mathcal{O}(1)$ time under per-UE parallelization. 
\item We implement the proposed estimator and scheduler in NetSim, a standard-compliant 5G emulator. Our emulation results show that MW-LC consistently outperforms conventional baselines such as Proportional Fair (PF) and Round Robin (RR) in AoI performance. 
We further evaluate the LC estimator and the MW-LC policy through MATLAB simulations. Our numerical results show that the LC estimator remains close to richer estimators in estimation accuracy, and that the resulting MW-LC policy achieves AoI performance close to that of richer estimator-based policies. 
\end{itemize}
More broadly, the estimator may be of independent interest to the community, enabling AoI-aware algorithms in different domains when the decision-maker is an intermediate entity that does not directly observe source-generation timestamps or destination freshness. 
\begin{figure}[t]
    \centering
    \includegraphics[width=\columnwidth,trim=0 0 0 0,clip]{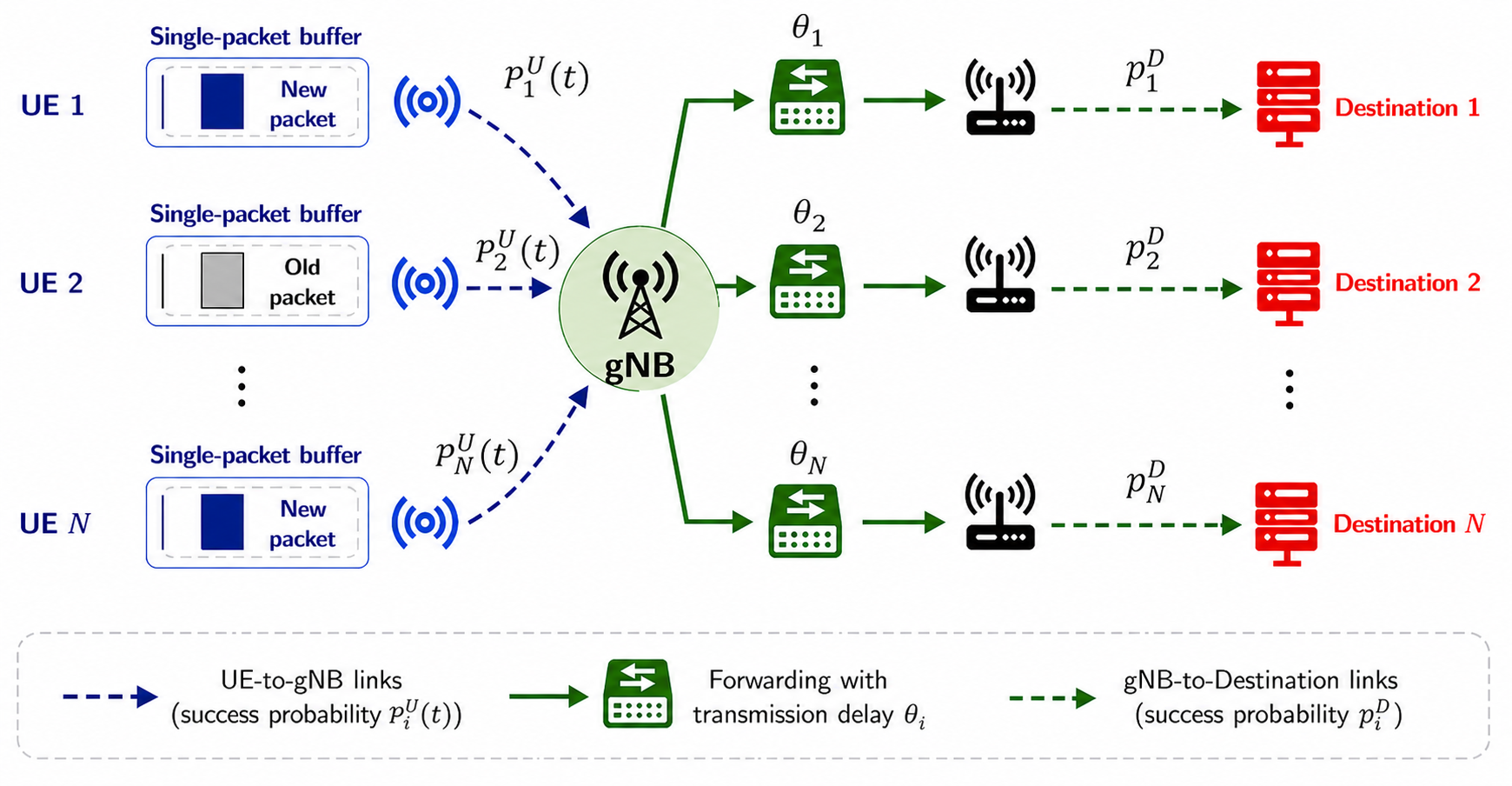}
\vspace{-1.2\baselineskip}
    \caption{Network with $N$ UE-destination pairs communicating through a 5G gNB. The gNB schedules $K$ uplink transmissions per slot. UE-to-gNB links have reliability $p_i^U(t)$, and successfully received packets are forwarded to the destinations over links with reliability $p_i^D$ and transmission delay $\theta_i$.}
    \label{fig:Systemmodel}
    \vspace{-1.5\baselineskip}
\end{figure}

\section{System Model}\label{sec:system}
Consider a network with $N$ UE sending time-sensitive uplink (UL) packets to $N$ destinations through a 5G gNB, as illustrated in Fig.~\ref{fig:Systemmodel}. Time is slotted, with slot index $t\in\{1,2,\ldots,T\}$. Each UE $i$ generates packets according to a renewal process. Let $g_i(t)=1$ indicate that UE $i$ generates a new packet at the beginning of slot $t$, and let $g_i(t)=0$ otherwise. Let the inter-generation period $X_i$ denote the number of slots between two consecutive packet generations from UE $i$, and define the average packet-generation rate as $\lambda_i := 1/\mathbb{E}[X_i].$ We assume\footnote{5G NR has periodic Scheduling Request (SR) opportunities through which UEs can inform the gNB that UL resources are needed. SR periodicity can be configured to be shorter/longer than a slot, with shorter periods resulting in higher control overhead. Multiple MAC/PHY conditions, including measurement gaps and SR prohibit timers, may prevent SR transmissions even if an SR opportunity exists. Our system model does not rely on real-time, perfect knowledge of UE packet generation. Instead, it relies on a long-term traffic statistic $\lambda_i$ that the gNB can estimate from scheduler-visible signaling and historical observations, with SR serving as one possible input.} that the gNB can estimate $\lambda_i$.

\vspace{0.5ex}
\noindent\textbf{System Time.}
Each UE maintains a \emph{single-packet queue}\footnote{Single-packet queues have been shown to be optimal in terms of AoI in many different scenarios~\cite{AoIsurvey}.} that stores only the freshest generated packet. Whenever a new packet is generated, it replaces the previously stored packet. If no new packet is generated, the stored packet remains available for (potential) repeated transmissions\footnote{The scheduler in Sec.~\ref{sec:MW} will attempt to avoid repeated transmission as they provide no freshness gain to the destination.}. Let $\gamma_i^U(t)$ denote the generation time of the single packet stored in UE $i$ at the beginning of slot $t$, and define its \emph{system time} as $a_i(t):=t-\gamma_i^U(t).$
The system time evolves as
\begin{equation}\label{eq:z_evol}
a_i(t+1)=
\begin{cases}
0, & \text{if } g_i(t)=1,\\
a_i(t)+1, & \text{otherwise.}
\end{cases}
\end{equation}

\vspace{0.5ex}
\noindent\textbf{Packet Transmission.}
At the beginning of each slot $t$, the gNB schedules at most $K$ UEs for UL transmission\footnote{Our system model can be extended to account for UL transmissions delayed by a constant number of slots. Furthermore, the scheduler in Sec.~\ref{sec:MW} can be easily modified to accommodate a time-varying $K$.}, where $K\leq N$. Let $u_i(t)\in\{0,1\}$ indicate whether UE $i$ is scheduled in slot $t$, with $\sum_{i=1}^N u_i(t)\leq K$.
A scheduled UE transmits its packet over a time-varying UE-to-gNB UL channel. Let $c_i^U(t)\in\{0,1\}$ denote the UE-to-gNB link success indicator in slot $t$, where $u_i(t)c_i^U(t)=1$ indicates that UE $i$'s packet is successfully received by the gNB. 
We model the UE-to-gNB UL channel using Sounding Reference Signal (SRS)-based measurements available at the gNB. In our model, the gNB obtains SRS-based measurements at the beginning of each frame\footnote{5G New Radio (NR) has 10\;ms frames, 1\;ms subframes, and slot duration determined by numerology. In Sec.~\ref{sec:results}, we assume slot duration of 1\;ms.} and uses them to estimate the UL SINR. These SRS-based measurements are mapped to a time-dependent UL link reliability, denoted by 
\(
p_i^U(t) \triangleq
\Pr\bigl(c_i^U(t)=1 \mid \mathrm{SRS}_i(t)\bigr).
\)
The SRS-based measurements and $p_i^U(t)$ are kept fixed within each frame and are available to the gNB when making slot-level scheduling decisions, while the realized link success indicator $c_i^U(t)$ becomes known only after the transmission attempt. 

Every packet received by the gNB is immediately forwarded to its corresponding destination through a heterogeneous network with long-term delivery reliability $p_i^D\in(0,1]$ and transmission delay $\theta_i\in\{0,1,\ldots\}$ slots. Let $c_i^D(t)\in\{0,1\}$ denote the gNB-to-destination link success indicator in slot $t$, where $u_i(t)c_i^U(t)c_i^D(t)=1$ indicates that UE $i$'s packet, transmitted in slot $t$, is successfully received by destination $i$ in slot $t+\theta_i$. We assume that $\{c_i^D(t)\}$ is independent over time slots and across UEs, with
$\Pr\bigl(c_i^D(t)=1\bigr)=p_i^D$.

\vspace{0.5ex}
\noindent\textbf{Age of Information (AoI).}
Let $\gamma_i^D(t)$ denote the generation time of the freshest packet successfully received by destination~$i$ by the beginning of slot $t$. The corresponding AoI is
$A_i(t):=t-\gamma_i^D(t).$
Its evolution is given by
\begin{equation}\label{eq:h_evol}
A_i(t+1)=
\begin{cases}
a_i(t-\theta_i)+\theta_i+1, & \text{if } \Lambda_i(t)=1,\\
A_i(t)+1, & \text{otherwise,}
\end{cases}
\end{equation}
where
$\Lambda_i(t):=u_i(t-\theta_i)c_i^U(t-\theta_i)c_i^D(t-\theta_i), \forall t>\theta_i.$
For simplicity, we assume $A_i(1)=1$ and $a_i(1)=0$ for all $i$.

For a non-anticipative scheduling policy $\pi\in\Pi$, the performance metric is the expected weighted sum AoI (EWSAoI):
\vspace{-1em}
\begin{equation}\label{eq:EWSAoI}
\mathbb{E}[J^\pi]
:=
\lim_{T\to\infty}\frac{1}{TN}\sum_{t=1}^T\sum_{i=1}^N \alpha_i\,\mathbb{E}[A_i^\pi(t)],
\end{equation}
where $\alpha_i>0$ is a priority weight for destination $i$.

\vspace{0.5ex}
\noindent\textbf{gNB Observation.}
The scheduler operates using gNB-side information from successful UL receptions. Upon each successful UL reception, the gNB determines whether the received packet is a repeated copy of the current most recent packet or a newly generated distinct packet\footnote{In the NetSim implementation, this distinction is made using the packet ID available in the simulator’s protocol stack. More generally, it requires packet-identity information exposed to the scheduler.}, and updates the stored packet-transmission times as described below. 

For each UE $i$, the gNB keeps track of the two most recently received distinct packets. Let $\tau_i^{\mathrm{cur}}(t)$ and $\bar{\tau}_i^{\mathrm{cur}}(t)$ denote the first and last slots, respectively, in which the \emph{current} most recent distinct packet was successfully received from UE $i$ by the beginning of slot $t$. Similarly, let $\tau_i^{\mathrm{pre}}(t)$ and $\bar{\tau}_i^{\mathrm{pre}}(t)$ denote the first and last slots, respectively, in which the \emph{previous} distinct packet was successfully received from UE $i$. Failed UL transmissions are not included in the estimator state, as they do not reveal the identity or generation time of the transmitted packet at the gNB~\cite{zhao2025optimizing}. History from older packets is not retained, as it is 
captured through the recursive estimator developed in Sec.~\ref{sec:LC-E}. 

Next, we describe an AoI estimator and scheduler that use knowledge of $N,K,p_i^D,\theta_i,\lambda_i$, and gNB observation
\vspace{-0.5ex}
\begin{equation}\label{eq:observation}
\mathbb{O}(t)\!:=\!
\left\{\!
\{p_i^U(t),\tau_i^{\mathrm{pre}}(t),\bar{\tau}_i^{\mathrm{pre}}(t),\tau_i^{\mathrm{cur}}(t),\bar{\tau}_i^{\mathrm{cur}}(t)\}_{i=1}^N
\!\right\}\!.
\end{equation}
We assume the gNB either knows or can estimate $p_i^D$, transmission delays $\theta_i$, and packet-generation rates $\lambda_i$ from network configuration, historical measurements, and scheduler-visible control information. In particular, $p_i^D$ and $\lambda_i$ are treated as long-term statistics rather than real-time observables, with $p_i^D\approx 1$ in many practical scenarios. Interestingly, we will see that timestamp estimation and the MW-LC do not depend on $\theta_i$, which only affects the AoI estimate $\hat{A}_i(t+\theta_i)$ in~\eqref{eq:hhat_LC}.  
\vspace{-1ex}

\section{Low-Complexity Estimation and Scheduling}\label{sec:LC-E}
\vspace{-1ex}

In this section, we develop a Low-Complexity (LC) Estimator for system time and AoI. 
Relative to \cite{zhao2025optimizing}, two main differences are: (i) assumption of knowledge about $\lambda_i$, as opposed to knowledge of the complete packet generation statistics, namely its probability mass function $\mathbb{P}(X_i=x),\forall i,x$; and (ii) a LC Estimator based on closed-form approximations, as opposed to high-complexity algorithms with nested ``for loops''. 
The resulting estimator trades exactness for tractability by replacing the full inter-generation distribution with a rate-based Bernoulli approximation.
Next, we leverage this restricted knowledge to derive approximate 
closed-form packet-generation probabilities, develop the LC Estimator and the Max-Weight-LC scheduling policy.
\subsection{Packet-Generation Probabilities}
For each UE $i$, the two most recent successful UL receptions determine the feasible interval in which the current packet may have been generated. Using a Bernoulli approximation parameterized by $\lambda_i$, we first define
\vspace{-1ex}
\begin{equation}\label{eq:lambda_D_LC}
\eta_i^{D}(t)
=
\frac{\lambda_i}
{1-(1-\lambda_i)^{\tau_i^{\mathrm{cur}}(t)-\bar{\tau}_i^{\mathrm{pre}}(t)}},
\end{equation}
which is the conditional generation probability associated with the interval
$[\bar{\tau}_i^{\mathrm{pre}}(t)+1,\tau_i^{\mathrm{cur}}(t)]$.

Let $q_i^{D}(\varphi)$ denote the conditional probability that the most recently received distinct packet of UE $i$ was generated in slot $\varphi$. Under this Bernoulli approximation, it follows that 
\begin{equation}\label{eq:g_D_LC}
q_i^{D}(\varphi)
=
\frac{
\eta_i^{D}(t)\,
(1-\eta_i^{D}(t))^{\tau_i^{\mathrm{cur}}(t)-\varphi}
}{
1-(1-\eta_i^{D}(t))^{\tau_i^{\mathrm{cur}}(t)-\bar{\tau}_i^{\mathrm{pre}}(t)}
}.
\end{equation}

Next, for the interval after the most recent successful reception of the current packet, we define
\begin{equation}\label{eq:lambda_S_LC}
\eta_i^{U}(t)
=
\frac{\lambda_i}
{1-(1-\lambda_i)^{t-\bar{\tau}_i^{\mathrm{cur}}(t)}},
\end{equation}
which is the conditional generation probability associated with the interval
$[\bar{\tau}_i^{\mathrm{cur}}(t)+1,t]$.

Similarly, let $q_i^{U}(\phi)$ denote the conditional probability that the packet currently stored at UE $i$ was generated in slot $\phi$. Under the Bernoulli approximation, it follows that 
\begin{equation}\label{eq:g_S_LC}
q_i^{U}\!(\!\phi\!)\!\!=\!\!
\begin{cases}
\!\eta_i^{D}\!(t)\bigl(1\!\!-\!\eta_i^{D}\!(t)\!\bigr)\!^{\tau_i^{\mathrm{cur}}\!(t)\!-\!\phi} \!C\!_i(t),\!\!
& \!\!\bar{\tau}_i^{\mathrm{pre}}\!(t)\!+\!1 \!\le\! \phi \!\le\! \tau_i^{\mathrm{cur}}\!(t),\!\\
\!\eta_i^{U}\!(t)\bigl(1\!-\!\eta_i^{U}(t)\!\bigr)\!^{t-\phi},
& \!\!\bar{\tau}_i^{\mathrm{cur}}\!(t)\!+\!1 \!\le\! \phi \!\le\! t,\\
\!0, & \!\!\text{otherwise.}
\end{cases}
\end{equation}
{\normalsize where} $C_i(t)\!\!:=\!\!\bigl(\!1\!-\!\eta_i^{U}(t)\!\bigr)^{t-\bar{\tau}_i^{\mathrm{cur}}(t)}
\!\!/\!\!\left(\!1\!\!-\!\!\bigl(\!1\!\!-\!\eta_i^{D}(t)\!\bigr)^{\tau_i^{\mathrm{cur}}(t)-\bar{\tau}_i^{\mathrm{pre}}(t)}\right)\!.$
\normalsize

The numerator in \eqref{eq:g_S_LC} corresponds to the event that the current packet is generated in slot $\varphi$ and no newer packet is generated before slot $t$, while the denominator normalizes this probability over the feasible generation window.

{
\setlength{\intextsep}{3pt plus 1pt minus 1pt}
\centering
\begin{minipage}{0.98\columnwidth}
\begin{algorithm}[H]
\caption{LC Estimator for $\hat{A}_i(t+\theta_i)$ and $\hat{a}_i(t)$}
\label{alg:LCE}
\begin{algorithmic}[1]
\STATE \textbf{Input:} Observation $\mathbb{O}(t)$ and estimate $\hat{\gamma}_i^D(t+\theta_i-1)$
\STATE Compute $\eta_i^{D}(t)$ and $\eta_i^{U}(t)$ using \eqref{eq:lambda_D_LC} and \eqref{eq:lambda_S_LC}
\STATE Compute $q_i^{D}(\varphi)$ and $q_i^{U}(\varphi)$ using \eqref{eq:g_D_LC} and \eqref{eq:g_S_LC}
\STATE Update $\hat{\gamma}_i^U(t)$ and $\hat{\gamma}_i^{D,\mathrm{cur}}(t)$ using \eqref{eq:tau_est_LC}
\STATE \textbf{If} $u_i(t)c_i^U(t)=1$ \textbf{then} Update $\hat{\gamma}_i^D(t+\theta_i)$ using \eqref{eq:tauD_rec1_LC}
\STATE \textbf{Else} Update $\hat{\gamma}_i^D(t+\theta_i)$ using \eqref{eq:tauD_rec0_LC}
\STATE Compute $\hat{A}_i(t+\theta_i)$ and $\hat{a}_i(t)$ using \eqref{eq:hhat_LC}
\STATE \textbf{Output:} $\hat{\gamma}_i^U(t)$, $\hat{\gamma}_i^D(t+\theta_i)$, $\hat{A}_i(t+\theta_i)$, and $\hat{a}_i(t)$
\end{algorithmic}
\end{algorithm}
\end{minipage}
}
\subsection{Low-Complexity Estimator}

Using the packet-generation probabilities in \eqref{eq:lambda_D_LC}--\eqref{eq:g_S_LC}, the gNB constructs approximate MMSE estimates of the UE timestamp and of the generation time of the most recently received distinct packet as
\begin{equation}\label{eq:tau_est_LC}
\hat{\gamma}_i^U(t)
=
\!\!\!\!\!\!\!\!\sum_{\varphi=\bar{\tau}_i^{\mathrm{pre}}(t)+1}^{t}\!\!\!\!\!\!\!\!
\varphi q_i^{U}(\varphi),\ \quad
\hat{\gamma}_i^{D,\mathrm{cur}}(t)
=
\!\!\!\!\!\!\!\!\sum_{\varphi=\bar{\tau}_i^{\mathrm{pre}}(t)+1}^{\tau_i^{\mathrm{cur}}(t)}\!\!\!\!\!\!\!\!
\varphi q_i^{D}(\varphi).
\end{equation}
To estimate the destination timestamp, the gNB recursively infers whether the most recently received distinct packet is delivered to the destination using the gNB-to-destination reliability $p_i^D$. 
If UE $i$ is successfully received by the gNB in slot $t$, then
\begin{equation}\label{eq:tauD_rec1_LC}
\hat{\gamma}_i^D(t+\theta_i)
=
(1-p_i^D)\hat{\gamma}_i^D(t+\theta_i-1)
+
p_i^D \hat{\gamma}_i^{D,\mathrm{cur}}(t).
\end{equation}
Otherwise,
\begin{equation}\label{eq:tauD_rec0_LC}
\hat{\gamma}_i^D(t+\theta_i)
=
\hat{\gamma}_i^D(t+\theta_i-1).
\end{equation}
Accordingly, the AoI and system-time estimates are given by
\begin{equation}\label{eq:hhat_LC}
\hat{A}_i(t+\theta_i)=t+\theta_i-\hat{\gamma}_i^D(t+\theta_i),\quad
\hat{a}_i(t)=t-\hat{\gamma}_i^U(t).
\end{equation}
Although $\hat{A}_i(t+\theta_i)$ is indexed by the forwarding delay,
the LC estimator itself updates the timestamp estimates
$\hat{\gamma}_i^U(t)$ and $\hat{\gamma}_i^D(t+\theta_i)$ using only
$\mathbb{O}(t)$, $\lambda_i$, and $p_i^D$. The delay $\theta_i$ is only
used when converting the destination timestamp estimate into the AoI
estimate in~\eqref{eq:hhat_LC}.
Notice that older packet-generation history does not need to be explicitly retained in the online LC update. Once slot $t$ begins, no new information about the generation time of earlier packets becomes available, and their effect on destination-side estimation has already been absorbed into the recursive state $\hat{\gamma}_i^D(t+\theta_i-1)$. Hence, for the purpose of LC estimation, the compressed observation in \eqref{eq:observation} is sufficient for updating the current UE timestamp and destination AoI estimates under the adopted recursive LC model. The resulting low-complexity estimator is summarized in Algorithm~\ref{alg:LCE}. 

The LC estimator has an average per-slot time complexity of $\mathcal{O}(N)$, with constant computation per UE. Its per-UE updates are separable and can be reduced to $\mathcal{O}(1)$ time under fully parallel execution. Therefore, incorporating the LC estimator does not increase the order of computational complexity of the Max-Weight scheduling policy introduced next.
\begin{figure}[t]
    \centering
    \includegraphics[width=0.9\linewidth]{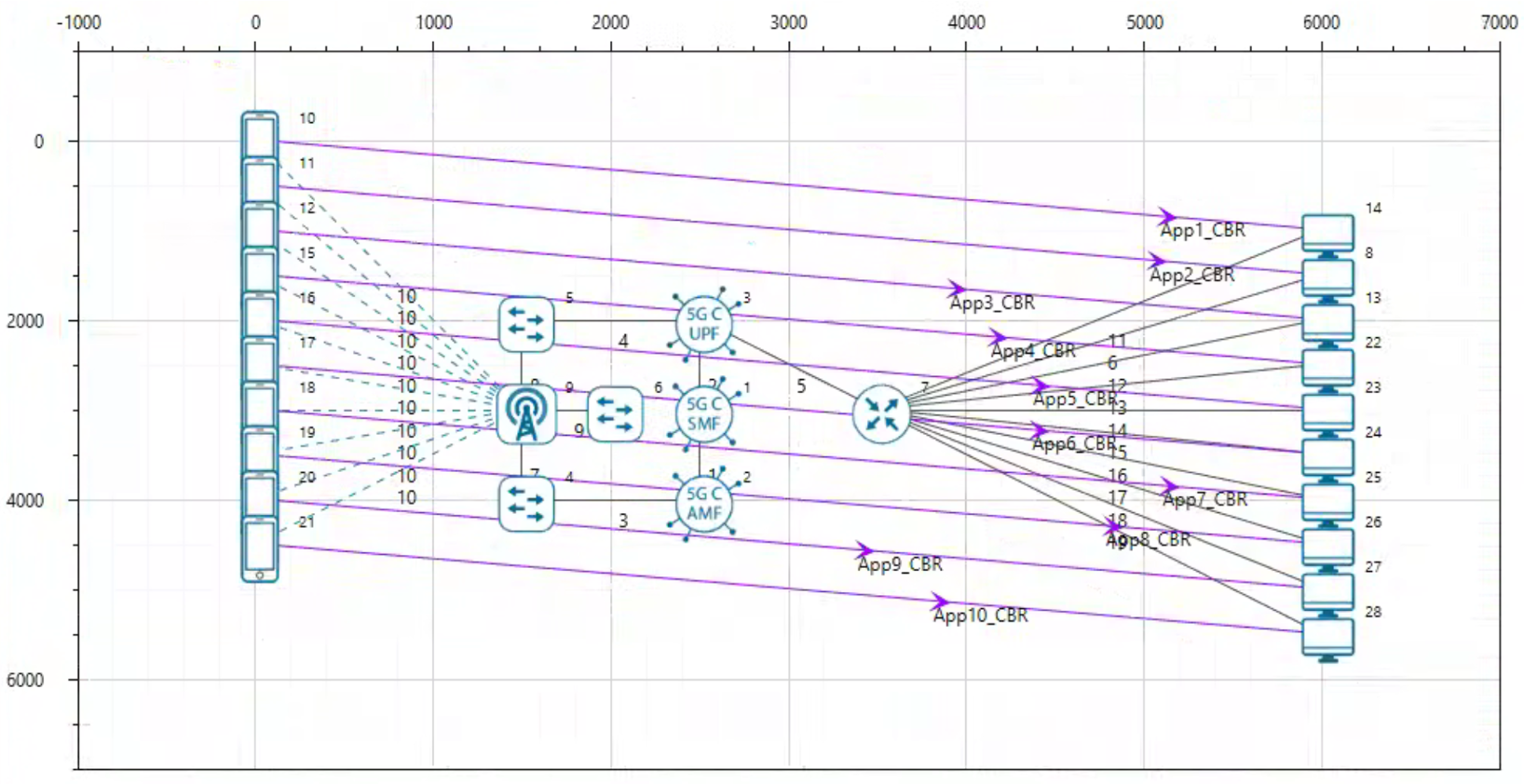}
    \caption{System Diagram of NetSim with $10$ UE-destination pairs.}
    \label{fig:netsim_architecture}\vspace{-0.5\baselineskip}
\end{figure}

\begin{table}[t]
\centering
\begin{minipage}{0.94\columnwidth}
\centering
\caption{Channel Model Setting}
\label{tab:channel_model_setting}
\renewcommand{\arraystretch}{1.12}
\begin{tabular}{p{0.46\linewidth} p{0.46\linewidth}}
\hline
\textbf{Setting} & \textbf{Value} \\
\hline
UE/gNB Height \& Antenna Count & 1.5\;m / 10\;m \& 1 / 2\\
UE Tx Power  & 23 dBm\\
CA Type \& CA Configuration & 2 Bands \& n78\\
DL:UL Ratio \& Numerology $\mu$ & 1:1 \& 0 (15\,kHz SCS, 1\,ms slot)\\
Channel BW \& Coherence Time & 40 MHz \& 10 ms\\
MCS Table & QAM256\\
Pathloss/Shadow Fading Model & 3GPP TR 38.901, Sec.~7.4.1 \\
Outdoor Scenario \& LOS/NLOS & URBAN\_MACRO \& LOS\\
Fading and Beamforming & RAYLEIGH\_WITH\_EIGEN \\
\hline
\end{tabular}
\end{minipage}
\vspace{-1.5\baselineskip}
\end{table}


\subsection{MW-LC Scheduling}\label{sec:MW}

Utilizing the LC estimates $\hat{A}_i(t+\theta_i)$ and $\hat{a}_i(t)$, we design an AoI-aware Max-Weight scheduling policy motivated by Lyapunov optimization~\cite{neely2022stochastic}. To account for heterogeneous transmission delays, we introduce the following Lyapunov function:
\vspace{-0.3em}
\begin{equation}\label{eq:lyapunov}
L(t)=\frac{1}{N}\sum_{i=1}^N \beta_i \hat{A}_i(t+\theta_i),
\end{equation}
where $\beta_i>0$ is a design parameter. Although $\hat{A}_i(t+\theta_i)$ refers to a future destination-side AoI estimate, it is deterministically computed from the current gNB observation and the LC estimator recursion.  
Since scheduling UE $i$ in slot $t$ may affect the destination AoI only at slot $t+\theta_i+1$, we consider the following one-slot Lyapunov drift:
$\Delta(\mathbb{O}(t))
:=
\mathbb{E}\!\left[L(t+1)-L(t)\mid \mathbb{O}(t)\right].$ 
%
Substituting the AoI evolution in~\eqref{eq:h_evol} into the drift yields
\begin{equation}\label{eq:drift_LC}
\Delta(\mathbb{O}(t))
=
\frac{1}{N}\sum_{i=1}^N \beta_i
-\frac{1}{N}\sum_{i=1}^N W_i(t)\,\mathbb{E}[u_i(t)\mid\mathbb{O}(t)],
\end{equation}
where
$W_i(t)
=
\beta_i\, p_i^U(t)\, p_i^D
\bigl(\hat{A}_i(t+\theta_i)-\hat{a}_i(t)-\theta_i\bigr)$ 
%
can be interpreted as the marginal drift-reduction score associated with scheduling UE $i$ under the current estimates and link reliabilities. 
Notice that $\hat{A}_i(t+\theta_i)-\hat{a}_i(t)-\theta_i
=
\hat{\gamma}_i^U(t)-\hat{\gamma}_i^D(t+\theta_i)$.
Hence, the explicit delay term cancels in the timestamp-difference form, and MW-LC does not require $\theta_i$ when computing the scheduling weights. To minimize the Lyapunov drift in~\eqref{eq:drift_LC}, the scheduler selects in each slot up to $K$ UEs with the largest positive weights $W_i(t)$, with ties broken arbitrarily.
We refer to this policy as \emph{MW-LC}. 

\begin{figure}[t]
    \centering
    \begin{minipage}{0.94\columnwidth}
        \centering
        \captionsetup[subfigure]{skip=0em}
        \begin{subfigure}[t]{0.95\linewidth}
            \centering
            \includegraphics[width=0.75\linewidth,trim=0 0 0 0,clip]{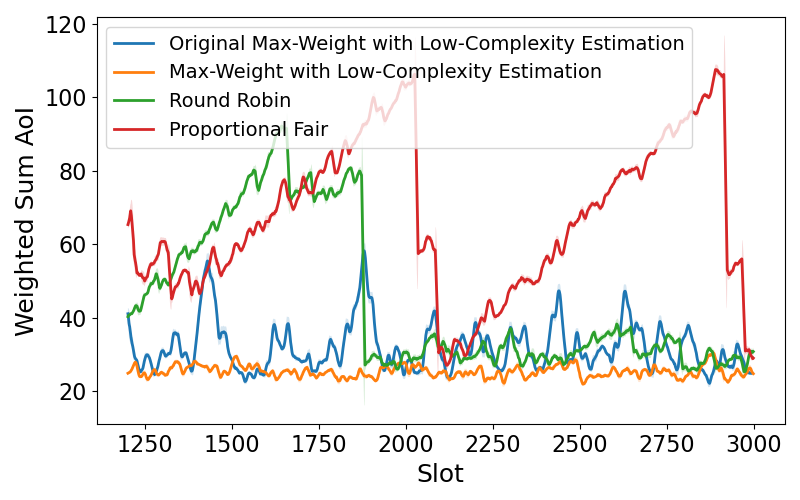}
            \caption{Weighted AoI sample path.}
            \label{fig:netsim_samplepath_trace}
        \end{subfigure}
        \vspace{-0.3ex}

        \begin{subfigure}[t]{0.95\linewidth}
            \centering
            \includegraphics[width=0.75\linewidth,trim=0 0 0 0,clip]{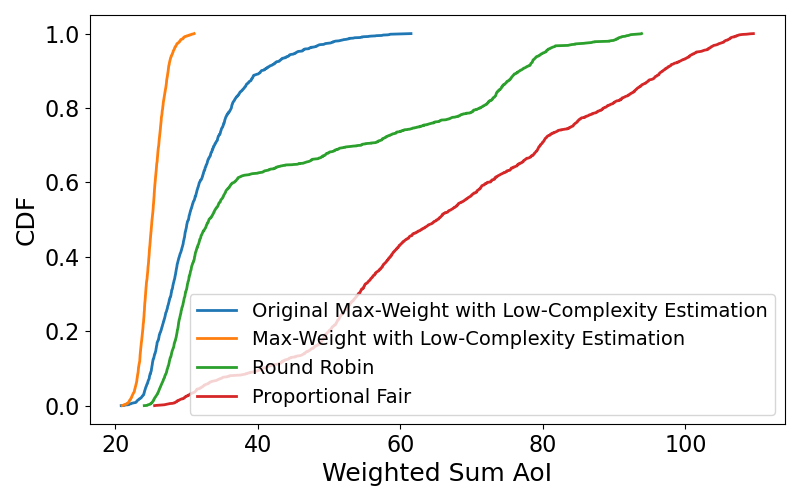}
            \caption{Empirical CDF of weighted AoI.}
            \label{fig:netsim_samplepath_cdf}
        \end{subfigure}

        \vspace{-0.3\baselineskip}
        \caption{NetSim 5G NR emulation results with $40$ UE-destination pairs and $K=2$ scheduled UL transmissions per slot. Bernoulli arrival rates are $\{0.05,0.2,0.5,1\}$ for UE groups $[1\!:\!10]$, $[11\!:\!20]$, $[21\!:\!30]$, and $[31\!:\!40]$, respectively.}
        \label{fig:netsim_samplepath}
    \end{minipage}
    \vspace{-1.5\baselineskip}
\end{figure}
\begin{figure}[t]
    \centering
    \begin{minipage}{0.96\columnwidth}
        \centering
        \includegraphics[width=0.7\linewidth]{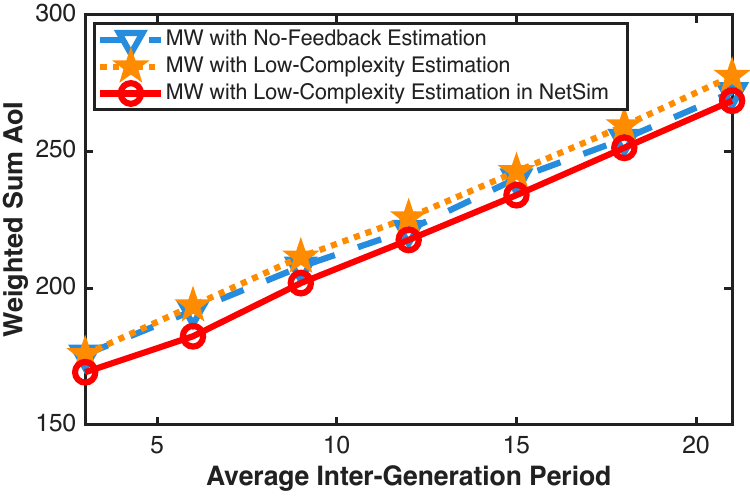}
        \caption{Comparison of NetSim LC estimator and MATLAB estimators under mixed periodic and Bernoulli arrivals, with $N=40$, $K=2$, and common expected inter-generation period $\mathbb{E}[X_i]$.}
        \label{fig:netsim_matlab_comparison}
    \end{minipage}
    \vspace{-1\baselineskip}
\end{figure}

\begin{figure}[t]
    \centering
    \begin{minipage}{0.96\columnwidth}
        \centering
        \includegraphics[width=0.7\linewidth]{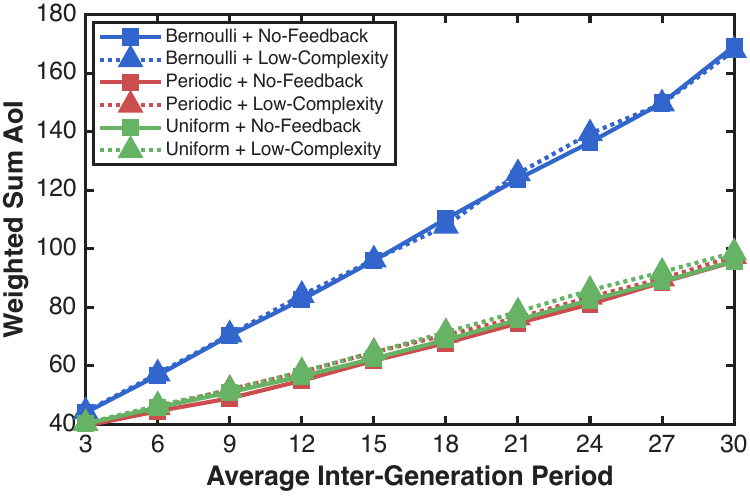}
        \caption{EWSAoI versus $\mathbb{E}[X_i]$ for $N=10$, $K=2$, $\boldsymbol{\alpha}=[1,2,3,4,5,2,4,6,8,10]$, NetSim-distributed uplink reliabilities $p_i^U(t)$, $p_i^D=0.8$, and $\theta_i=5$, under uniform, Bernoulli, and periodic packet generation.}
        \label{fig:performance_varying_interarrival}
    \end{minipage}
    \vspace{-1.5\baselineskip}
\end{figure}
\section{Emulation and Simulation Results}\label{sec:results}

This section evaluates the proposed framework from two complementary perspectives. We first implement the scheduler in a 5G NR emulator using the LC estimator-based policy and only scheduler-visible information at the gNB. We then implement the MW-LC scheduler in MATLAB, compare it with the NetSim results, and compare it with the higher-complexity estimator-based policy\footnote{As shown in Fig.~\ref{fig:runtime_vs_users_combined}, the high-complexity estimator incurs substantial per-slot runtime and is therefore too costly for direct implementation in the NetSim 5G NR protocol stack. We thus evaluate this baseline only in MATLAB. The MATLAB code is available at~\cite{AoI_LCEstimator_code}.} from~\cite{zhao2025optimizing}.
\subsection{NetSim Emulation}
We implement the scheduler in NetSim~\cite{netsim}, a discrete-event simulator that models the 5G NR protocol stack, including the application, MAC, and PHY layers. Packet generation occurs at the application layer, UL transmissions are scheduled at the MAC layer, and decoding outcomes together with SRS-based UL channel measurements are produced at the PHY layer. In line with our model assumptions, the NetSim gNB can only observe network configuration parameters including $N,K,p_i^U(t),p_i^D,\lambda_i$, transmission outcomes, and SRS-based UL channel-state information, from which it executes the MW-LC scheduler without direct access to UE/destination timestamps. In implementation, the mapping from SRS-based UL measurements to the reliability $p_i^U(t)$ is estimated offline via Monte Carlo simulation under the same channel model. 

Figure~\ref{fig:netsim_architecture} illustrates the network implemented in NetSim. The wireless channel and PHY-layer parameters follow the 3GPP-compliant configuration provided by the simulator~\cite{netsim} and are summarized in Table~\ref{tab:channel_model_setting}. 


\noindent\textbf{NetSim Results.}
We compare against the original estimator-based Max-Weight policy (MW-O) in~\cite{zhao2025optimizing}, as well as Round-Robin (RR) and Proportional Fairness (PF)~\cite{kelly1998rate}. MW-O assumes an i.i.d. UE-to-gNB channel and uses long-term UL reliabilities. All policies are evaluated under identical channel realizations and traffic processes. We do not include~\cite{WiSwarm,AoI5G,li2024aequitas} as direct numerical baselines because they are formulated for different scheduling objectives and gNB observability models, which may conflate the comparison. In particular, our focus is the setting in which the gNB does not directly observe UE-side timestamps or destination freshness and must estimate them from gNB-visible observations. 

We emulate a network with $N=40$ UEs and schedule $K=2$ UL transmissions per slot. The UEs are divided into four traffic classes with Bernoulli arrival rates $0.05$, $0.2$, $0.5$, and $1$, respectively, and heterogeneous priority weights $\alpha_i$ across users. Channel reliabilities and transmission delays are jointly determined by the user locations, the channel configuration in Table~\ref{tab:channel_model_setting}, and the protocol settings of the emulator.

Fig.~\ref{fig:netsim_samplepath} shows the weighted Sum AoI sample path and CDF obtained from the NetSim emulation. 
The gNB executes the LC estimator using only protocol-level observations, and the true AoI at the destinations is never revealed to the scheduler. This shows that the proposed observation model and scheduling policy can be realized within the NetSim protocol stack, under the observability assumptions described in Sec.~\ref{sec:system}, without revealing explicit AoI values to the scheduler.

Among all implemented policies, MW-LC achieves the lowest AoI in the tested NetSim scenarios. Conventional channel-driven schedulers such as PF can repeatedly favor users in good channel states and may ignore stale users for extended periods, leading to large AoI build-ups. Moreover, because MW-O relies on long-term channel reliabilities, it may repeatedly select a user whose instantaneous channel remains poor within a transmission frame, whereas the SRS-aware MW-LC policy mitigates this effect by incorporating frame-level UL reliability into the scheduling weight. In the scenario of Fig.~\ref{fig:netsim_samplepath}, this corresponds to average AoI reductions of $63\%$, $44\%$, and $21\%$ relative to PF, RR, and MW-O, respectively.

\subsection{MATLAB Simulation}
\begin{figure*}[t]
    \centering
    \begin{minipage}[t]{0.65\textwidth}
        \centering
        \begin{subfigure}[t]{0.49\textwidth}
            \centering
            \includegraphics[width=\textwidth,trim=0 0 0 0,clip]{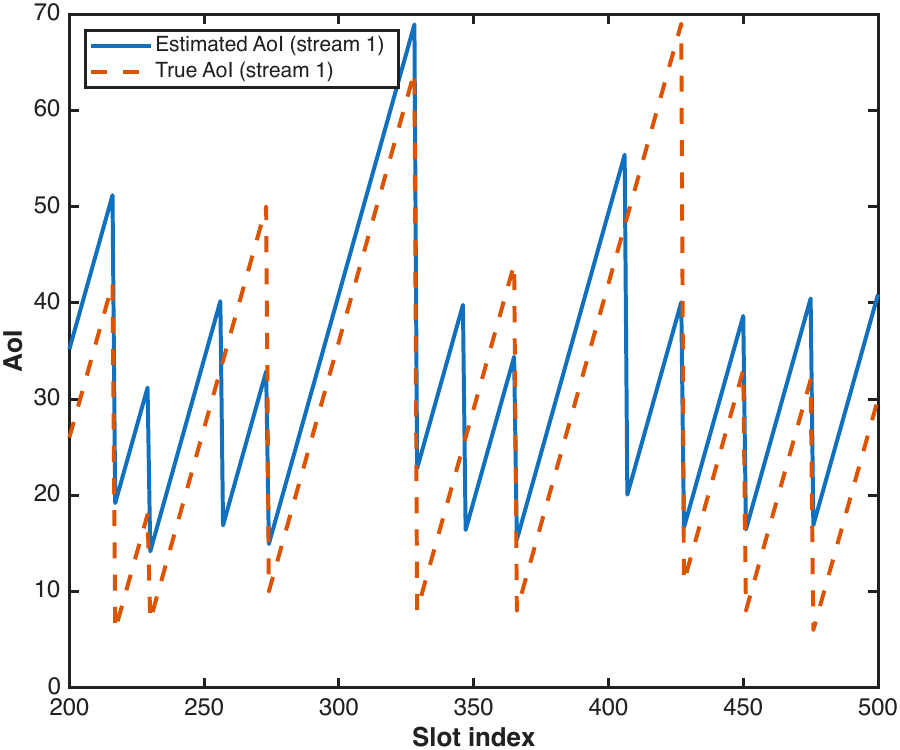}
            \caption{AoI sample path of Stream~1 with $p_i^D\!=\!0.8,\ \forall i$.}
            \label{fig:mismatch_samplepath_pdi_08}
        \end{subfigure}
        \hfill
        \begin{subfigure}[t]{0.49\textwidth}
            \centering
            \includegraphics[width=\textwidth,trim=0 0 0 0,clip]{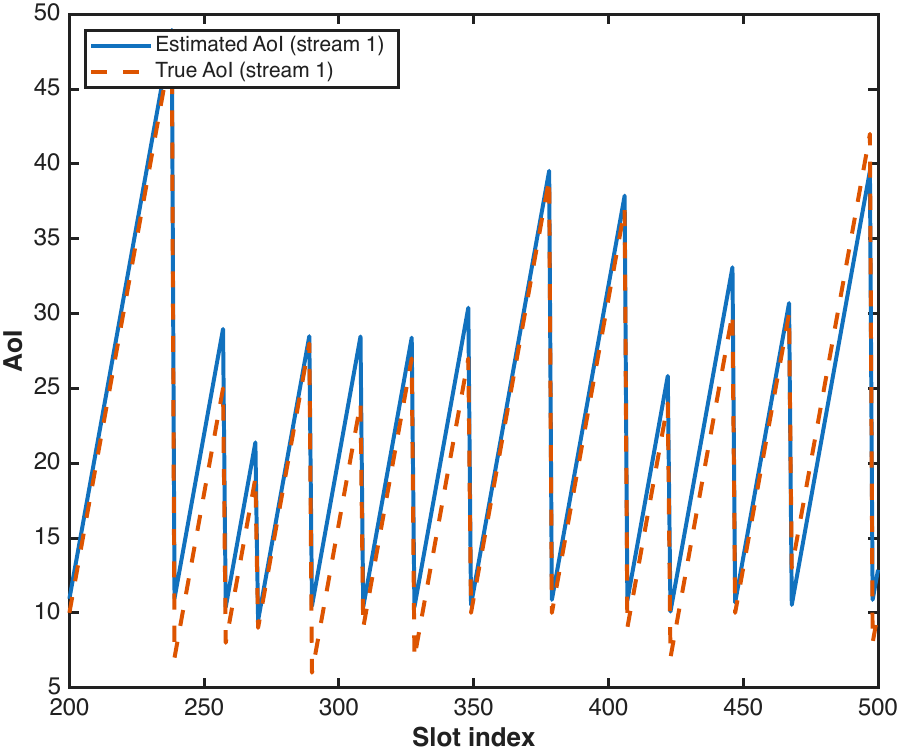}
            \caption{AoI sample path of Stream~1 with $p_i^D=1,\ \forall i$.}
            \label{fig:mismatch_samplepath_pdi_1}
        \end{subfigure}
        \captionof{figure}{True and estimated AoI sample paths for a representative stream under the LC estimator. Parameters: $N=10$, $K=2$, $\boldsymbol{\alpha}=[1,2,3,4,5,2,4,6,8,10]$, uplink reliabilities $p_i^U(t)$ distributed as in the first 10 UEs of the NetSim configuration, $\theta_i=5$, and $X_i\sim U[4,8]$. The destination-link reliability is $p_i^D\!=\!0.8$ in (a) and $p_i^D\!=\!1$ in (b).}
        \label{fig:mismatch_samplepath_combined}
    \end{minipage}
    \hfill
    \begin{minipage}[t]{0.33\textwidth}
        \centering
        \includegraphics[width=0.98\textwidth]{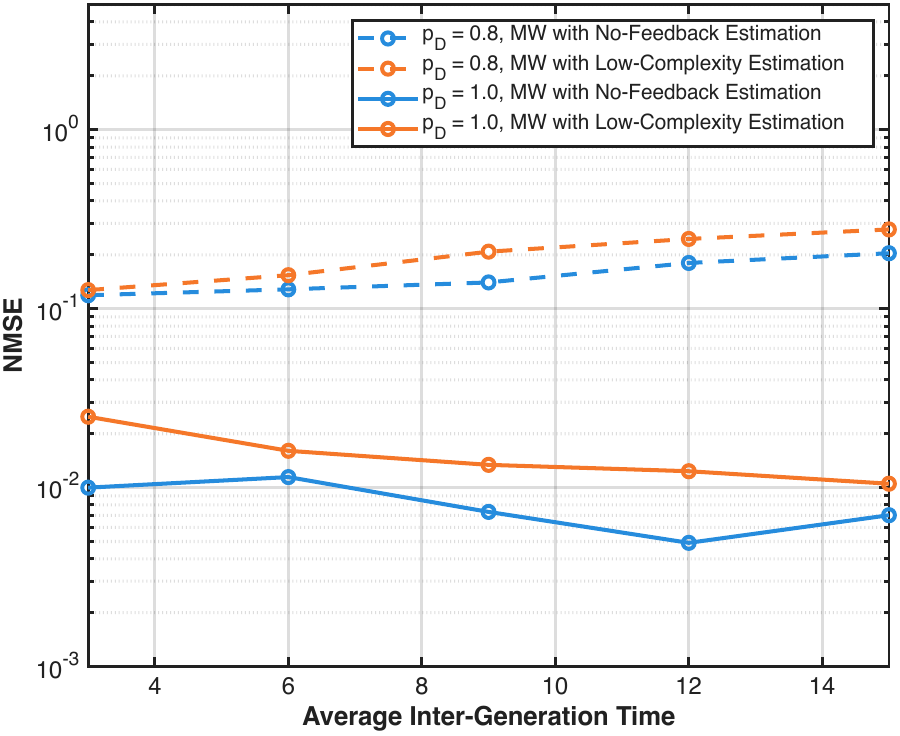}
        \captionof{figure}{Estimator NMSE versus the expected inter-generation period $\mathbb{E}[X_i]$ for $N\!=\!10$, $K\!\!=\!\!2$, \!$\boldsymbol{\alpha}\!=\![1\!,2\!,3,\!4,\!5,\!2$,$4$,$6$,$8$,$10]$, $p_i^U\!(t)\!$ distributed as in the first 10 UEs of the NetSim configuration, \!and $\theta_i\!\!=\!\!1$. Each UE follows $X_i\!\sim\! U[0.8\,\mathbb{E}[X_i],\,1.2\,\mathbb{E}[X_i]]$. The destination-link reliabilities are $p_i^D\!=\!\!0.8$ and $p_i^D\!\!=\!\!1$.}
        \label{fig:nmse_vs_EX}
    \end{minipage}
    \vspace{-1.5\baselineskip}
\end{figure*}
We next use MATLAB simulations to compare MW-LC against the Max-Weight with no-feedback Estimation (MW-EnF)~\cite{zhao2025optimizing}, whose richer estimator exploits the probability mass function of the inter-generation period $\mathbb{P}(X_i=x),\forall i,x$. 
The performance of both policies is averaged over $10$ independent runs with horizon $T=10^5$ slots.

Fig.~\ref{fig:netsim_matlab_comparison} compares the LC estimator implemented in the NetSim 5G NR emulator with the MATLAB-based estimators under mixed periodic and Bernoulli arrivals. 
The gap between MW-LC and the richer estimators remains small across the tested inter-generation periods, suggesting that the Bernoulli approximation preserves much of the information most relevant for scheduling.

Figs.~\ref{fig:performance_varying_interarrival} evaluates the policies under uniform, Bernoulli, and periodic packet-generation processes, respectively. 
Across all these generation processes MW-EnF and MW-LC exhibit very similar performance.
Figs.~\ref{fig:mismatch_samplepath_combined} further compare the true AoI and the LC-estimated AoI for a representative stream. When $p_i^D=1$, the estimated AoI closely matches the true AoI because every forwarded packet is delivered. 
When $p_i^D=0.8$, a mismatch appears because the gNB cannot directly observe destination receptions and must instead infer them recursively. Nevertheless, the two trajectories remain well aligned in their drop instants and relative magnitudes, indicating that the LC estimator preserves the urgency information needed for scheduling. This observation is further supported by Fig.~\ref{fig:nmse_vs_EX}, where the estimator NMSE remains moderate while the AoI performance of MW-LC stays close to that of MW-EnF.

\vspace{-1.5ex}
\section{Final Remarks}
\label{sec:conclusion}

This paper studied AoI-aware scheduling for 5G cellular under limited freshness observability and slot-level runtime constraints. We developed a low-complexity estimator (based on gNB-visible observations) and a MW-LC scheduling policy that leverages SRS-based UL CSI. We implemented the estimator and scheduler in NetSim, a 5G NR emulator with a standards-compatible protocol stack, and showed that MW-LC outperforms 5G baselines. We used MATLAB simulations to evaluate both the estimator and scheduler under a broader set of traffic and network conditions.  
Our current design assumes access to long-term traffic parameters, packet-identity information needed for distinct-packet tracking, and a Bernoulli approximation of packet generation. 
Interesting directions for future work include implementing MW-LC on Software Defined Radios and within the OpenAirInterface software stack, as well as applying our low-complexity estimator framework to other domains in which the decision-maker is an intermediate entity (e.g., a network cache, federated learning server, network controller, or digital twin) that does not directly observe source-generation timestamps or destination freshness.




%

\vspace{-1ex}

\bibliographystyle{IEEEtran}
\bibliography{reference}
\ifCLASSOPTIONcaptionsoff
  \newpage
\fi

\begin{IEEEbiography}[{%
\includegraphics[
    height=1.25in,
    clip,
    keepaspectratio,
    trim=10 0 20 0
]{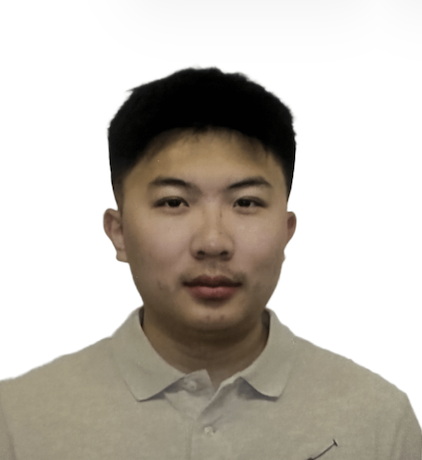}}]{Zhuoyi Zhao(Graduate Student Member, IEEE)}
 received the B.E. degree in Information Engineering from Beijing Jiaotong University, Beijing, China, in 2023, and the M.S. degree in Electrical Engineering from Northwestern University, Evanston, IL, USA, in 2025. He is currently a Graduate Research Assistant with the Department of Electrical and Computer Engineering, University of Toronto, Toronto, ON, Canada. His research is on optimization and control in communication networks, online learning, federated learning, and multi-agent systems. 
\end{IEEEbiography}
\vskip -1\baselineskip plus -1fil
\begin{IEEEbiography}[{%
\includegraphics[
    height=1.2in,
    clip,
    keepaspectratio,
    trim=20 0 20 0
]{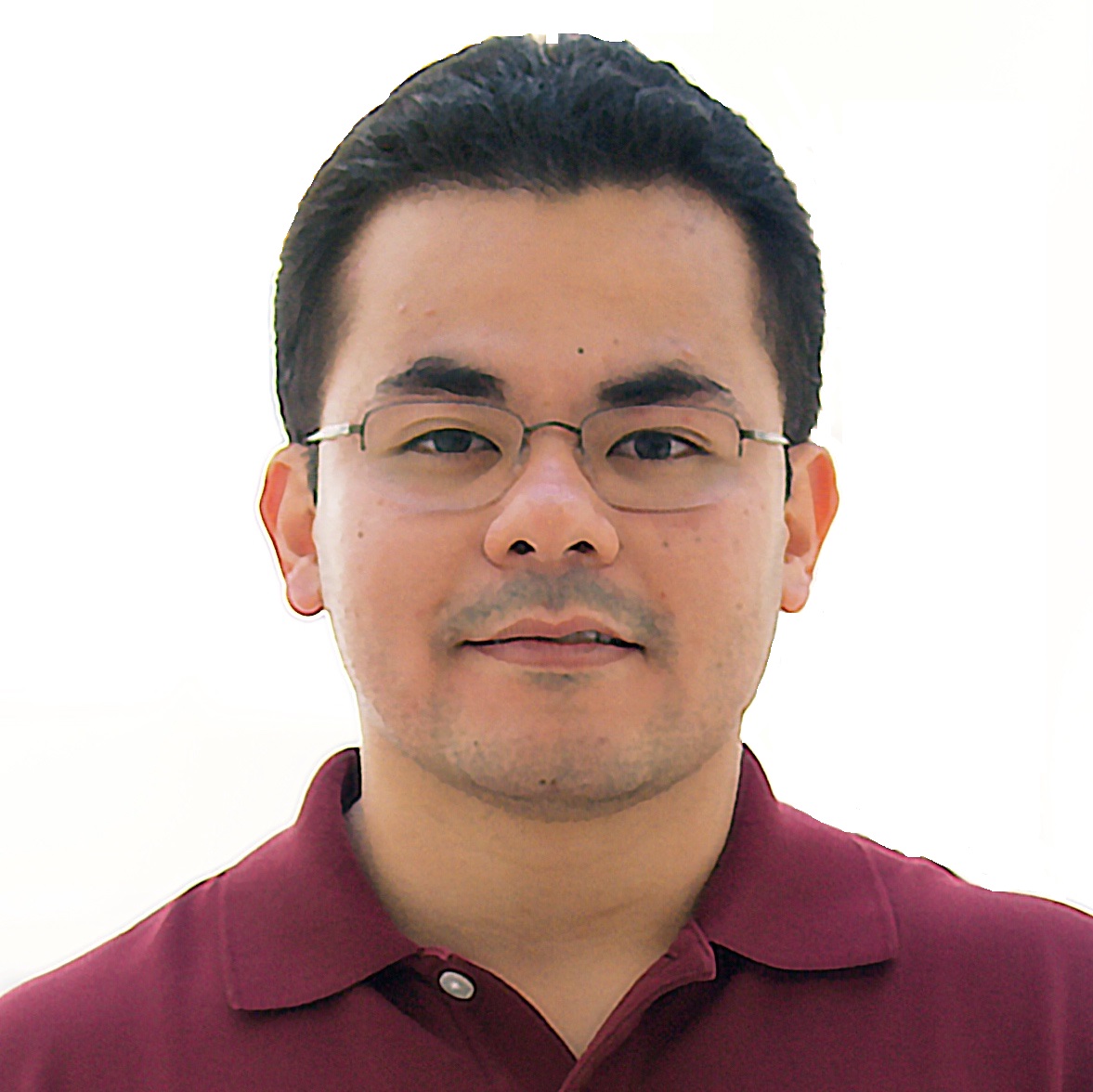}}]{Igor Kadota (Senior Member, IEEE)}
received the B.S. degree in electronic engineering and the S.M. degree in telecommunications from the Aeronautics Institute of Technology (ITA), Brazil, in 2010 and 2013, and the S.M. and Ph.D. degrees in communication networks from the Massachusetts Institute of Technology (MIT), in 2016 and 2020, respectively. He was a Postdoctoral Research Scientist at Columbia University from 2020 to 2023. He is currently an Assistant Professor with the Department of Electrical and Computer Engineering, Northwestern University. His research is on modeling, analysis, optimization, and implementation of emerging communication networks, with emphasis on wireless networks and time-sensitive traffic.

He was a recipient of several research, teaching, and mentoring awards, including the Best Paper Award at IEEE INFOCOM 2018, the Best Paper Award Finalist at ACM MobiHoc 2019, the Best Student Paper Award at WiOpt 2024 and at WiOpt 2025, the MIT School of Engineering Graduate Student Extraordinary Teaching and Mentoring Award of 2020, and the 2019--2020 Thomas G.~Stockham Jr.~Fellowship.
\end{IEEEbiography}



\end{document}